\documentclass[twocolumn]{aastex631}

\usepackage{graphicx}	
\usepackage{amsmath}	
\usepackage{amssymb}	

\shorttitle{Discussions on GLEAM-X J1627}
\shortauthors{H. Tong}
\graphicspath{{./}{figures/}}

\begin{document}

\title{Discussions on the nature of GLEAM-X J162759.5$-$523504.3}

\correspondingauthor{H. Tong}
\email{tonghao@gzhu.edu.cn}

\author[0000-0001-7120-4076]{H. Tong}
\affiliation{School of Physics and Materials Science, Guangzhou University, Guangzhou 510006, China}









\begin{abstract}
The nature of the long period radio transient GLEAM-X J162759.5$-$523504.3 (GLEAM-X J1627 for short) is discussed. We try to understand both its radio emission and pulsation in the neutron star scenario, as an alternative to the white dwarf model. We think that: (1) From the radio emission point of view, GLEAM-X J1627 can be a radio-loud magnetar. (2) From the rotational evolution point of view, GLEAM-X J1627 is unlikely to be an isolated magnetar. (3) The 1091s period is unlikely to be the precession period. (4) GLEAM-X J1627 may be a radio-loud magnetar spin-down by a fallback disk. (5) The pulsar death line is modified due to the presence of a fallback disk or a twisted magnetic field. In both cases, a higher maximum acceleration potential can be obtained. This may explain why GLEAM-X J1627 is still radio active with such a long pulsation period. (6) General constraint on the neutron star magnetic field and initial disk mass are given analytically. Possible ways to discriminate between different modelings are also discussed.
\end{abstract}

\keywords{stars: magnetar -- pulsars: general -- pulsars: individual (GLEAM-X J162759.5$-$523504.3)}


\section{Introduction}

Recently, a transient radio source with a possible period of 1091 seconds (about 18 minutes) is reported (Hurley-Walker et al. 2022) . It is thought to be a long period radio emitting magnetar in the discovery paper. We would like to comment on this possibility and give our discussions about the nature of GLEAM-X J162759.5$-$523504.3 (here after GLEAM-X J1627 for short).

After more than 50 years, we know a lot about pulsars and magnetars. For their rotational behaviors, the slowest radio pulsar at present is PSR J0250$+$5854 with a period of 23.5 seconds (Tan et al. 2018). It may be spin-down by magnetospheric processes or involving magnetic field decay (Kou et al. 2019). Possible precession signal in pulsars (Stairs et al. 2000, Ashton et al. 2017; with period about 1000 days) and magnetars (Makishima et al. 2014, 2019, with period about 0.5 days) are also found. The precession may be free precession (Ashton et al. 2017; Makishima et al. 2019) or forced precession due to the presence of a fallback disk (Qiao et al. 2003).
Possible period of 16 and 159 days is also reported in two fast radio bursts (The CHIME/FRB Collaboration 2020; Rajwade et al. 2020). This long period may be due to binary origin or forced precession due to a fallback disk (Lyutikov et al. 2020; Yang \& Zou 2020; Ioka \& Zhang 2020; Tong et al. 2020).

The central compact object inside the supernova remnant RCW 103 is confirmed to be a magnetar (D'Ai et al. 2016; Rea et al. 2016). Its 6.6 hours period may be the rotational period of the central magnetar (De Luca et al. 2006; D'Ai et al. 2016; Rea et al. 2016). It may be spin-down by the presence of a fallback disk (Tong et al. 2016). Different combination of magnetic field strength and fallback disk mass may explain the behavior of normal magnetars with period about 10 s and the magnetar with 6.6 hour period (Tong et al. 2016). At present, the magnetar inside RCW 103 is the slowest isolated neutron star.


Compared with previous observations, GLEAM-X J1627's period of 1091 seconds is not very surprising. It is long compared with that of normal pulsars and normal magnetars. However, it is much shorter compared with that of RCW 103 magnetar and that of possible precession signal in pulsars, magnetars, and fast radio bursts etc.

By applying previous experiences in pulsars and magnetars, we think that:  GLEAM-X J1627 may be radio-loud magnetar spin-down by a fallback disk. From figure 1 in Tong et al. (2016), a fallback disk accreting neutron star can naturally result in periods about $10^3 \ \rm s$, which we think is the case of GLEAM-X J1627. Therefore, GLEAM-X J1627 (with a period about $1091 \ \rm s$) may be an intermediate object between normal magnetars (with period about $10 \ \rm s$) and the magnetar inside RCW 103 (with a period of 6.6 hours).

\subsection{Summary of observations}

From Hurley-Walker et al. (2022), GLEAM-X J1627 has a flux of (5-40) Jy, is observed in the frequency range (72-231) MHz, is at a distance about $1.3 \ \rm kpc$, has a brightness temperature $\sim 10^{16} \ \rm K$ (which requires a coherent emission process), has a period of $1091 \ \rm s$, has an upper limit on period derivative $\dot{P} < 1.2\times 10^{-9} \ \rm s \ s^{-1}$, and has an upper limit of X-ray luminosity $L_x < 10^{32} \ \rm erg \ s^{-1}$.

From the observational flux and distance, the isotropic radio luminosity is estimated to be:
\begin{eqnarray}
  &&L_{\rm iso} \sim f \times \nu \times 4\pi d^2\\
   &&= 4\times 10^{30} \ {\rm erg \ s^{-1}} \left(\frac{d}{1.3\ \rm kpc}\right)^2 \frac{f}{10\ \rm Jy}
  \frac{\nu}{200 \ \rm MHz},
\end{eqnarray}
where $f$ is the typical observed flux, $\nu$ is the observational frequency, $d$ is the source distance. More exact calculation of the radio luminosity will require the beam radius, duty cycle and spectra information (Szary et al. 2014). From the observed period and period derivative, a lower limit on the characteristic age is:
\begin{equation}
  \tau_c = \frac{P}{2\dot{P}} > 1.4 \times 10^4 \ {\rm yr}.
\end{equation}
An upper limit on the characteristic magnetic field is (surface dipole magnetic field strength at the equator):
\begin{eqnarray}
  B_c &=&3.2 \times 10^{19} \times \sqrt{P \dot{P}} I_{45}^{1/2} R_6^{-3} \ \rm G \\
  &<& 3.7 \times 10^{16} I_{45}^{1/2} R_6^{-3} \ \rm G,
\end{eqnarray}
where $I_{45}$ is moment of inertial in units of $10^{45} \ \rm g \ cm^2$, $R_6$ is the star radius in units of $10^6 \ \rm cm$. An upper limit on the rotational energy loss rate is:
\begin{equation}
\label{eqn_Edot}
  \dot{E} = \frac{4\pi^2 I \dot{P}}{P^3} < 3.6 \times 10^{28} I_{45} \ \rm erg \ s^{-1}.
\end{equation}

For a typical neutron star with $I_{45} \approx 1$, $R_6 \approx 1$, the radio luminosity of GLEAM-X J1627 is larger than the neutron star's rotational energy loss rate. Possible beaming may soften this problem. Detailed calculations for GLEAM-X J1627 can be found in Erkut (2022). However, for normal pulsars, their radio luminosity is always much smaller than the rotational energy loss rate. Therefore, even considering the effect of beaming, GLEAM-X J1627 is very different from normal radio pulsars.
Therefore, the problem of GLEAM-X J1627 is always two fold (Hurley-Walker et al. 2022): (1) what's the energy budget for the radio emission, (2) what's the origin for its long pulsation period? Any modeling for GLEAM-X J1627 should address these two problems simultaneously.

As can be seen from eq.(\ref{eqn_Edot}), a white dwarf will have a much higher rotational energy loss rate compared with that of the pulsar case, because the moment of inertia of the white dwarf is much larger than that of the neutron star. Therefore, a white dwarf model can easily account for the energy budget and long pulsation period (Loeb \& Maoz 2022; Katz 2022). Coincidentally, the white dwarf model was also proposed as an alternative model for magnetars observations (Paczynski 1990; Malheiro et al. 2012).

As an alternative to the white dwarf model, we will try to provide an understanding of GLEAM-X J1627 in the neutron star scenario.

\section{On the nature of GLEAM-X J1627}

\subsection{From the radio emission point of view, GLEAM-X J1627 can be a radio-loud magnetar}

The mean flux (averaged over the whole period) of radio pulsar is order of $\rm mJy$. While their peak flux is order $\rm Jy$ (Lyne \& Graham-Smith 2012). For the radio emitting magnetar XTE J1810$-$197, at $3.3 \ \rm kpc$, its peak luminosity is about $10 \ \rm Jy$ (Camilo et al. 2006). The third radio emitting magnetar PSR J1622$-$4950 is radio-loud while in X-ray quiescence with $L_X \le 10^{33} \ \rm erg \ s^{-1}$ (Levin et al. 2010; Anderson et al. 2012). Therefore, both the radio luminosity and low X-ray luminosity of GLEAM-X J1627 may be similar to a radio-loud magnetar in X-ray quiescence, as also noted by the discovery paper (Hurley-Walker et al. 2022). In this scenario, the radio emission of GLEAM-X J1627 is powered by the magnetic energy of a magnetar. Future discovery of long period sources with high X-ray luminosity and X-ray outburst (i.e. radio-loud magnetar not in X-ray quiescence) will give direct support for the magnetar scenario, similar to the confirmation of magnetar inside RCW 103 (D'Ai et al. 2016; Rea et al. 2016).

\subsection{From the rotational evolution point of view, GLEAM-X J1627 is unlikely to be an isolated magnetar}

For both radio pulsars and radio-loud magnetars, they all lie above a fiducial pulsar death line on the $P-\dot{P}$ diagram (Ruderman \& Sutherland 1975; Zhou et al. 2017). This fiducial death line can be defined as: the maximum acceleration potential across the polar cap region equals $10^{12} \ \rm V$ (Ruderman \& Sutherland 1975; Zhou et al. 2017):
\begin{equation}\label{eqn_Phimax}
  \Phi_{\rm max} = \frac{B_{\rm p} R^3 \Omega^2}{2c^2} \equiv 10^{12} \ \rm V,
\end{equation}
where $\Omega=2\pi/R$ is the angular velocity of the neutron star, and $B_{\rm p} = 6.4\times 10^{19} \sqrt{P \dot{P}}$ is the surface magnetic field at the pole region, which is two times the commonly reported equatorial surface magnetic field (Lyne \& Graham-Smith 2012). Although the definition of this pulsar death line involves acceleration potential across the polar cap, it is just a fiducial death line when plotted on the $P-\dot{P}$ diagram of pulsars (Zhou et al. 2017). For a $1091 \ \rm s$ pulsar to lie above this fiducial pulsar death line, the required period derivative is: $\dot{P} \ge 10^{-8} \ \rm s \ s^{-1}$. However, the observational upper limit on the period derivative is: $\dot{P} \le 1.2 \times 10^{-9}$. Therefore, GLEAM-X J1627 is unlikely to lie above the pulsar death line. One way to overcome this difficulty is to involve physical definitions of pulsar death lines (Zhang et al. 2000).

Furthermore, according to the observational upper limit on period derivative, the required surface magnetic field is: $B_c \le 3.7 \times 10^{16} \ \rm G$ and characteristic age: $\tau_c \ge 1.4 \times 10^4 \ \rm yr$ (see the above summary of observations). However, for a neutron star with a magnetic field of $10^{16} \ \rm G$, its persistent X-ray luminosity will also be relatively high (Vigano et al. 2013). This is in contradiction with the upper limit on X-ray luminosity $L_x \le 10^{32} \ \rm erg \ s^{-1}$. For normal magnetars, the typical magnetic field is $\sim 10^{15} \ \rm G$, with luminosity $10^{33}- 10^{35} \ \rm erg \ s^{-1}$ (Coti Zelati et al. 2018). If the true period derivative of GLEAM-X J1627 is two orders of magnitude smaller: $\dot{P} \sim 10^{-11}$, the requirement of magnetic field strength will be softened (down to $10^{15} \ \rm G$). However, the required timescale to spin-down to the long rotational period will be: $\tau_c \sim 10^6 \ \rm yr$. The magnetic field strength will decay significantly during this long timescale (Rea et al. 2010; Vigano et al. 2013; Kou et al. 2019). With only magnetospheric braking mechanism, it is hard to spin-down a neutron star
to a period of $1091 \ \rm s$.

In conclusion, from the rotational evolution point view, GLEAM-X J1627 is unlikely to lie above the pulsar death line, and unlikely to be spin-down to its present long period.

\subsection{The 1091s period is unlikely to be the precession period}

For normal pulsars, the typical rotational period is: $P \sim 0.1 \ \rm s$. If the neutron star is deformed under
the influence of an internal toroidal magnetic field of $B_t \sim 10^{16} \ \rm G$, the ellipticity of the neutron star is (Makishima et al. 2019; Tong et al. 2020):
\begin{equation}
  \varepsilon \sim 10^{-4} \left( \frac{B_t}{10^{16} \ \rm G} \right)^2.
\end{equation}
The corresponding period of free precession is: $P_{\rm precession} = P/\varepsilon \sim 10^3 \ \rm s$. This may explain the pulsation period of GLEAM-X J1627 (Eksi \& Sasmaz 2022).

However, as discussed in the discovery paper (Hurley-Walker et al. 2022), the period of GLEAM-X J1627 is very accurate: $\sigma_P/P < 5\times 10^{-7}$. Therefore, exact periodic mechanism are preferred, i.e. rotational or orbital period (Hurley-Walker et al. 2022). While free precession may only result in quasi-periodicity of neutron stars (Staris et al. 2000; Ashton et al. 2017). The reason for a quasi-periodicity may be two fold: (1) the fluid core of the neutron star will result in damping of the oscillation (Shaham 1977; Sedrakian et al. 1999); (2) The spin-down torque due to the magnetosphere, both near field and far field, will cause the precession to be torqued precession instead of free precession (Gao et al. 2020). Furthermore, as stated above, a magnetic strength of $10^{16} \ \rm G$ is hard to reconcile with the low X-ray luminosity (Vigano et al. 2013).

A neutron star may also experience forced precession in the presence of a fallback disk (Qiao et al. 2003; Tong et al. 2020). However, the corresponding precession period is about several days or tens of days (eq.(7) and (10) in Tong et al. 2020). The 1000 days period in PSR B1828-11, and 16-day/159-day period in fast radio burst may be due to forced precession by a fallback disk. However, the 18 minutes period of GLEAM-X J1627 is too short to be explained by the forced precession.

In conclusion, the $1091 \ \rm s$ period of GLEAM-X J1627 is unlikely to be due to precession, either free precession or forced precession.

\subsection{GLEAM-X J1627 as a radio-loud magnetar spin-down by a fallback disk}
\label{sect_FBdisk}

Normal magnetars have typical period about $10 \ \rm s$ (Olausen \& Kaspi 2014). Two normal magnetars (4U 0142+61 and 1E 2259+586) may have passive fallback disks (Wang et al. 2006; Kaplan et al. 2009). The central compact object inside supernova remnant RCW 103 has a pulsation period about $6.6$ hours (De Luca et al. 2006; D'Ai et al. 2016; Rea et al. 2016). It may be spin-down by the presence of a fallback disk (Tong et al. 2016). A magnetar+fallback disk system may provide a unified explanation for normal magnetars, magnetars with fallback disks, and the magnetar inside RCW 103. Then it is natural that some source with period lying between $10 \ \rm s$ and $6.6$ hours can be seen.

\begin{figure}[htbp!]
  \centering
  \includegraphics[width=0.45\textwidth]{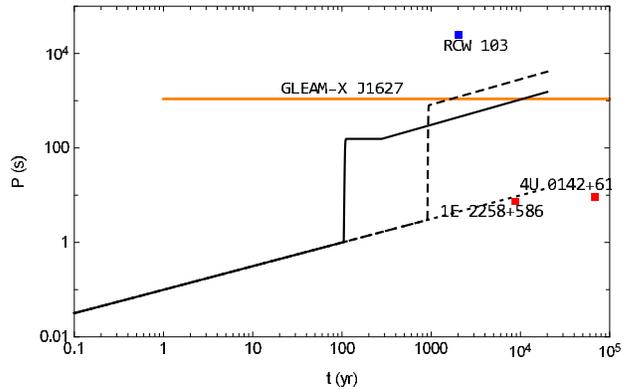}\\
  \caption{Rotational evolution of magnetars in the presence of a fallback disk. The magnetic field is chosen as $4\times 10^{14} \ \rm G$. The solid, dashed, and dotted lines are for initial disk mass of $10^{-3} \ \rm M_{\odot}$, $10^{-4} \ \rm M_{\odot}$, $10^{-5} \ \rm M_{\odot}$, respectively. GLEAM-X J1627 is represented by a line, since its age is unknown. The two magnetars with possible fallback disks 4U 0142+61 and 1E 2259+586, the magnetar inside RCW 103 with a pulsation period of $6.6$ hours are also shown. The calculations are stopped at an age of $2\times 10^4 \ \rm yr$, which is the typical age of fallback disks.}
  \label{fig_gleamx}
\end{figure}

Applying the modeling in Tong et al. (2016), the calculations for GLEAM-X J1627 in shown in figure \ref{fig_gleamx}. The major input is a high magnetic field neutron star, spin-down by a self-similar fallback disk, under a unified spin-up and spin-down accretion torque. In figure \ref{fig_gleamx}, the neutron star magnetic field is chosen $4\times 10^{14} \ \rm G$, ten times the critical magnetic field, similar to the radio-loud magnetar PSR J1622-4950 (Levin et al. 2010). Three typical initial disk mass are shown: $10^{-3} \ \rm M_{\odot}$, $10^{-4} \ \rm M_{\odot}$, $10^{-5} \ \rm M_{\odot}$. An initial disk mass about $10^{-3}- 10^{-4} \ \rm M_{\odot}$ may explain the $1091 \ \rm s$ period of GLEAM-X J1627. When the disk mass is small, e.g. $10^{-5} \ \rm M_{\odot}$, the disk can not enter into the neutron star magnetosphere.  This is because a smaller disk mass will result in a smaller mass accretion rate and a larger accretion magnetosphere radius. When the magnetospheric radius is larger than the neutron star light cylinder radius, the disk will not interact with the neutron star and it will be a passive fallback disk. This may corresponds to the fallback disk in magnetars 4U 0142+61 and 1E 2259+586 (Wang et al. 2006; Kaplan et al. 2009), as pointed in Tong et al. (2016). From figure 1, it can be seen that, there is a large parameter space (magnetic field, initial disk mass) for the rotational evolution of GLEAM-X J1627. The calculations in Ronchi et al. (2022) is similar to Tong et al. (2016) and the calculations here. Ronchi et al. (2022) is highly numerical, while the calculations here are to a large extent analytical. Numerical calculations are employed mainly in the final step.

Therefore, from our previous experiences for the $6.6$ hour magnetar inside RCW 103, GLEAM-X J1627 may be a magnetar spin-down by a fallback disk. Its $1091 \ \rm s$ lies between that of normal magnetars and the magnetar inside RCW 103. Combining radio emission and timing requirement, GLEAM-X J1627 may be a radio-loud magnetar spin-down by a fallback disk.

\subsection{Modification of pulsar death line for long period radio pulsars}
\label{sect_death_line}

\begin{figure}[htbp!]
  \centering
  \includegraphics[width=0.45\textwidth]{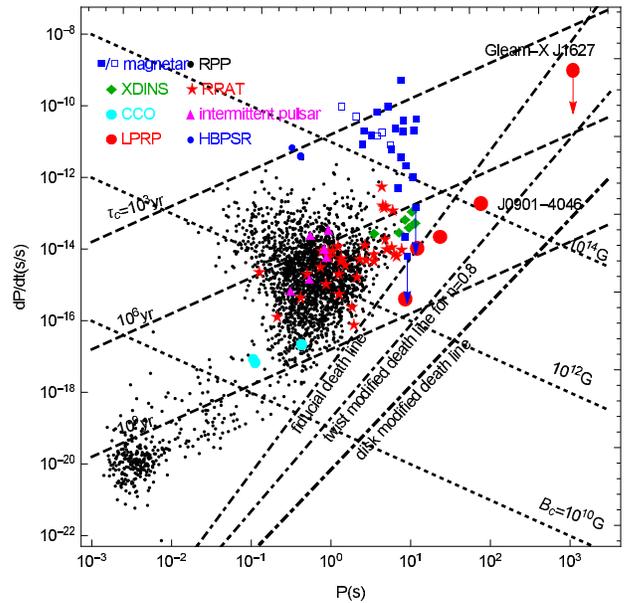}\\
  \caption{Distribution of long period radio pulsars (red circles) on the $P$-$\dot{P}$ diagram of pulsars. The fiducial pulsar death line, the death line modified by the fallback disk, the death line for a twisted dipole field ($n=0.8$) are also shown. It can be seen that a fallback disk or a twisted magnetosphere may help to explain why long period radio pulsars can still have radio emissions. The $P$-$\dot{P}$ diagram of various pulsar-like objects are updated from figure 2 in Kou et al. (2019).}
  \label{fig_LPRP}
\end{figure}

For a large scale dipole magnetic field, the potential drop across the polar cap with angular extent $\theta_{\rm pc}$ is (Ruderman \& Sutherland 1975; Tong 2016):
\begin{equation}\label{eqn_phi_max_theta}
  \Phi_{\rm max} = \frac{B_{\rm p} R^2 \Omega}{2c} \sin^2\theta_{\rm pc}.
\end{equation}
The dipole field line equation is: $r=r_e \sin^2\theta$, where $r_e$ is the maximum radial extent of the field lines. When the light cylinder is chosen as the maximum radio extent, the corresponding maximum acceleration potential is the commonly reported case, shown in eq.(\ref{eqn_Phimax}). This is the fiducially pulsar death line (Zhou et al. 2017), shown in figure \ref{fig_LPRP}. According to this fiducial pulsar death line, GLEAM-X J1627 already lies below the death line. The question is: how can a 1091s neutron star still have radio emissions?

There are two possible physical effects that may help overcome this difficulty: an active fallback disk or a twisted magnetosphere. If the fallback disk around GLEAM-X J1627 is still active, then both effects can contribute. If the fallback disk is no longer active, and GLEAM-X J1627 can now be treated as an isolated magnetar, then only the latter effect is possible. Whether the fallback disk is active or not is not known at present (Ronchi et al. 2022; Gencali et al. 2022; Rea et al. 2022).

(1) \textit{Death line modified by a fallback disk.} In the disk accretion case, the magnetospheric radius defines the maximum radio extent of the closed field lines (Ghosh \& Lamb 1979; Shapiro \& Teukolsky 1983). In accretion equilibrium, the corotation radius is equal to the magnetospheric radius (Fu \& Li 2013). Therefore, the corotation radius defines the maximum radial extent of the closed field lines. Then the maximum acceleration potential across the polar cap is:
\begin{equation}
  \Phi_{\rm max,disk} = \frac{B_{\rm p} R^3 \Omega^2}{2c^2} \frac{R_{\rm lc}}{R_{\rm co}},
\end{equation}
where $R_{\rm lc}$ and $R_{\rm co}$ is the light cylinder radius and corotation radius, respectively. In the presence of fallback disk accretion, the potential drop across the polar cap is enhanced by a factor $R_{\rm lc}/R_{\rm co}$. And the definition of pulsar death line will be modified by the presence of a fallback disk: $\Phi_{\rm max,disk} \equiv 10^{12} \ \rm V$.

For GLEAM-X J1627 with a pulsation period of $1091\ \rm s$, the potential drop is: $\Phi_{\rm max,disk} = 1.6 \times 10^{11} B_{14} R_6^3 \ \rm V$. For magnetic field several times of $10^{14} \ \rm G$, the potential drop can be near the critical value of $10^{12} \ \rm V$. Therefore, considering the presence of a fallback disk, GLEAM-X J1627 may still have a high enough potential to acceleration particles and emit radio emissions. Its transient nature may because it lies near the pulsar death line.

Normally, the radio emission will be ceased during accretion, as demonstrated by the transitional millisecond radio pulsars (Papitto \& de Martino 2022). For accreting neutron stars, the accretion may only occur in an annular region of the polar cap (Ghosh \& Lamb 1978; Frank et al. 2002). This is due to a finite width of the boundary layer at the magnetospheric radius. Therefore, the core region of the polar cap may still permits particle acceleration and radio emission. This possibility is originally discussed in the fallback disk model for the observations of anomalous X-ray pulsars and soft gamma-ray repeaters (Ertan et al. 2009; Trumper et al. 2010), as an alternative to the magnetar model. The difference between fallback accreting neutron stars and normal accreting neutron stars may be that: the neutron star is spinning down (instead of spin-up) due to a decreasing mass accretion rate of the fallback disk (Chatterjee et al. 2000; Alpar 2001).

(2) \textit{Death line for a twisted magnetic field.} Magnetars may have twisted magnetic field compared with that of normal pulsars (Thompson et a. 2002; Beloborodov 2009; Pavan et al. 2009). A twisted magnetic field will result in a larger polar cap (Tong 2019). This will also result in a larger potential drop across the polar cap.

For a twisted dipole field, the radial dependence of the magnetic field is: $B(r) \propto r^{-(2+n)}$ (Wolfson 1995), where $n=1$ corresponds to the dipole case, $n=0$ corresponds to the split monopole case, $0<n<1$ corresponds to a twisted dipole case. Due to inflation of the field line in the radial direction of a twisted dipole field, more field lines will become open and a larger polar cap will be expected (Tong 2019). According to eq.(12) in Tong (2019), the polar cap for a twisted dipole field is:
\begin{equation}
  \sin^2\theta_{\rm pc} \approx \left( \frac{R}{R_{\rm lc}} \right)^n.
\end{equation}
Again, $n=1$ corresponds to the dipole case. According to eq.(\ref{eqn_phi_max_theta}), the maximum acceleration potential for a twisted dipole field is:
\begin{equation}
  \Phi_{\rm max,twist} = \frac{B_{\rm p} R^3 \Omega^2}{2c^2} \left( \frac{R_{\rm lc}}{R} \right)^{1-n}.
\end{equation}
For $n=1$, the maximum acceleration potential returns to the dipole case. The death line in the case of a twisted dipole field may be defined as: $\Phi_{\rm max,twist} \equiv 10^{12} \ \rm V$.

The distribution of long period radio pulsars on the $P$-$\dot{P}$ diagram is shown in figure \ref{fig_LPRP}. The five long period radio pulsars include: GLEAM-X J1627 (Hurley-Walker et al. 2022), the recently discovered $76 \ \rm s$ radio pulsars PSR J0901$-$4046 (Caleb et al. 2022), along with the three previously known long period radio pulsars (Tan et al. 2018). The fiducial pulsar death line, the death line modified by the presence of a fallback disk, and the death line for a twisted dipole field (for $n=0.8$) are shown. The presence of a fallback disk or a twisted magnetic field will lower the position of death line on the $P$-$\dot{P}$ diagram. These two effects may explain why GLEAM-X J1627 and other long period radio pulsars can still have radio emissions.

\subsection{Constraint on the magnetic field and disk mass}

The major input for a neutron star+fallback disk system are the neutron star's magnetic field strength and the initial disk mass (Chatterjee et a. 2000; Alpar 2001; Wang et al. 2006; Tong et al. 2016). The light cylinder radius, magnetospheric radius, and mass accretion rate (for a self-similar fallback disk) can all be expressed analytically. Therefore, some analytical constraint on the magnetic field and initial disk mass can be obtained.

The neutron star with a fallback disk will (1) firstly be spin-down under its own magnetic dipole field, (2) enter into the propeller regime and be quickly spin-down, (3) acquire accretion equilibrium with the disk (Tong et al. 2016). In order for the fallback disk to enter into the neutron star's light cylinder, the magnetospheric radius should be smaller than the light cylinder radius:
\begin{equation}
  R_{\rm m} (t) \le R_{\rm lc} (t),
\end{equation}
where the two radii both evolves with time. The light cylinder radius is: $R_{\rm lc} = P(t) c/2\pi \propto \mu \ t^{1/2}$, where $\mu$ is the magnetic dipole moment, the period evolution with time can be approximated by the dipole braking (eq.(11) in Tong 2016; eq.(5.18) in Lyne \& Graham-Smith 2012) before it interact with the fallback disk. The magnetospheric radius: $R_{\rm m} \propto \mu^{4/7}\ t^{5/14}$ (Tong et al. 2016, see footnote 7 there for the definition of magnetospheric radius and eq.(4) there for the accretion rate as a function of time). The lower limit on the magnetic field strength in order for the disk to enter into the neutron star light cylinder is:
\begin{equation}\label{eqn_Blower_1}
  B \ge 4\times 10^{13} \left( \frac{M_{\rm d,0}}{10^{-3} \ \rm M_{\odot}} \right)^{-2/3} \left( \frac{t}{10^4 \ \rm yr} \right)^{-1/3} \ \rm G,
\end{equation}
where $M_{\rm d,0}$ is the initial disk mass, $t$ is the typical age of a fallback disk. The initial mass of the fallback disk may be in the range $(10^{-6},\ 0.1) \ \rm M_{\odot}$ (Michel 1988; Chevalier 1989; Wang et al. 2006; Perna et al. 2014).

The neutron star will be quickly spin-down during the ejector phase and acquire accretion equilibrium with the fallback disk. When the magnetospheric radius is equal to the corotation radius, the corresponding period is defined as the equilibrium period (eq.(9) in Tong et al. 2016):
\begin{equation}
\label{eqn_Peq}
  P_{\rm eq} = 915 B_{15}^{6/7} \dot{M}_{\rm acc,17}^{-3/7} \ {\rm s} \propto B^{6/7} t^{3\alpha/7},
\end{equation}
where $\alpha=5/4$ for a Kramers opacity dominated disk. In order to spin-down the neutron star to the observed pulsation period in less than the typical age $t$, it is required that: $P_{\rm eq} \ge P_{\rm obs}$, where $P_{\rm obs}$ is the observational pulsation period. The lower limit on the magnetic field is:
\begin{equation}\label{eqn_Blower_2}
\scriptsize
  B \ge 3.7\times 10^{14} \left( \frac{M_{\rm d,0}}{10^{-3} \ \rm M_{\odot}} \right)^{1/2}
  \left( \frac{t}{10^4 \ \rm yr} \right)^{-5/8} \left( \frac{P_{\rm obs}}{10^3 \ \rm s}  \right)^{7/6} \ \rm G.
\end{equation}

\begin{figure}[htbp!]
  \centering
  \includegraphics[width=0.45\textwidth]{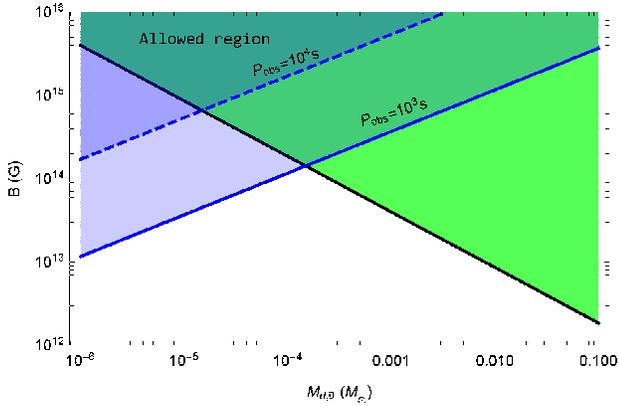}\\
  \caption{Constraint on magnetic field and initial disk mass. The black solid line is the lower limit for magnetic field, in order for the disk to enter into the neutron star light cylinder (eq.\ref{eqn_Blower_1}). The blue solid line is the lower limit for the magnetic field in order to spin-down it to the observed pulsation period (eq.\ref{eqn_Blower_2}), for $P_{\rm obs} =10^3 \ \rm s$. The blue dashed line is for $P_{\rm obs} = 10^4 \ \rm s$. The typical age of the fallback disk is chosen as $10^4 \ \rm yr$.}
  \label{fig_gBMd}
\end{figure}

The above two constraint on the magnetic field as a function of initial disk mass is plotted in figure \ref{fig_gBMd}. As can be seen from figure \ref{fig_gBMd}, for longer pulsation period $P_{\rm obs}$ the required magnetic field will be higher. This is why magnetars are always employed for long period pulsars, both isolated and accreting ones. For a self-similar fallback disk, the initial disk mass is proportional to the initial mass accretion rate (eq.(2) in Tong et al. 2016). Therefore, figure \ref{fig_gBMd} here and figure 5 in Ronchi et al. (2022) are consistent with each other. Figure 5 in Ronchi et al. (2022) are specific calculations for GLEAM-X J1627, while figure \ref{fig_gBMd} here are general constraint on fallback accreting neutron stars.

From figure \ref{fig_gBMd}, more analytical constraints can be obtained. In the allowed region, there exist a lower limit on the magnetic field. Combining eq.(\ref{eqn_Blower_1}) and eq.(\ref{eqn_Blower_2}), the lower limit on the magnetic field is:
\begin{equation}
  B \ge 1.4 \times 10^{14} \left( \frac{P_{\rm obs}}{10^3 \ \rm s} \right)^{2/3} \left( \frac{t}{10^4 \ \rm yr} \right)^{-1/2} \ \rm G.
\end{equation}
For a longer period, the required magnetic field is also higher. The intersection point between the two line moves up-left for a longer period.

The initial mass of the fallback disk may may have a lower limit about $10^{-6} \ \rm M_{\odot}$ (Michel 1988; Chevalier 1989; Wang et al. 2006; Perna et al. 2014).
The neutron star magnetic field may have an upper limit about $10^{16} \ \rm G$ (Duncan \& Thompson 1992; Olausen \& Kaspi 2014). Combining these two constraints, there exists an upper limit on the period of fallback accreting neutron stars. From eq.(\ref{eqn_Blower_2}) and considering the limit on disk mass and magnetic field, the upper limit on the neutron star period is:
\begin{equation}
\scriptsize
  P_{\rm obs} \le 3\times 10^5 \left( \frac{B}{10^{16} \ \rm G} \right)^{6/7} \left( \frac{M_{\rm d,0}}{10^{-6} \ \rm M_{\odot}} \right)^{-3/7} \left( \frac{t}{10^4 \ \rm yr} \right)^{15/28}  \ \rm s,
\end{equation}
which is about several days for a disk age about $10^4$ years. These analytical constrains can be applied to more long period radio pulsars in the future.

\section{Discussions}

As an alternative to the white dwarf model, the long period radio transient GLEAM-X J1627 is modeled as a radio-loud magnetar spin-down by a fallback disk. Future observations may help to discriminate between different modelings, shown in below.

\subsection{Comparison with other modelings}

For the radio emission and long pulsation period of GLEAM-X J1627, these two aspects can be explained naturally in the white dwarf model (Loeb \& Maoz 2022; Katz 2022). The physics of white dwarf pulsars may be similar to that of pulsars (Goldreich \& Julian 1969; Ruderman \& Sutherland 1975). Optical observations may help to discriminate between the white dwarf and the neutron star model (Hurley-Walker et al. 2022; Rea et al. 2022). Since neutron stars can sustain smaller period, future observations of more radio transients with smaller period may also help to clarify whether they are neutron stars or white dwarfs.

It can not be excluded that the long pulsation period of GLEAM-X J1627 is due to precession (Eksi \& Sasmaz 2022). However, the exactness of period may favor rotational or orbital period (Hueley-Walker et al. 2022). Our previous experiences in pulsars, magnetars and fast radio bursts tell us that precession may only result in quasi-periodicity (Stairs et al. 2000; Makishima et al. 2019; Tong et al. 2020). Future period observation of more sources may tell us whether their period is exact or quasi-periodic. Furthermore, if two periods can be found in one source (one spin period+one modulation period), then the precession or orbital origin may be preferred.

A normal neutron star (with $B \sim 10^{12} \ \rm G$) with a fallback disk may also explain the long period of GLEAM-X J1627 (Gencali et al. 2022). The accretion equilibrium period depends on both the magnetic field and accretion rate (see eq.(\ref{eqn_Peq})): $P_{\rm eq} \propto B^{6/7} \dot{M}^{-3/7}$. For a low magnetic field, a low mass accretion rate is required to produce the same period. Then the required initial disk mass should be smaller and typical age of the system should be larger. This is consistent with the quantitative result of Gencali et al. (2022). The difference between Gencali et al. (2022) amd the calculation here (section \ref{sect_FBdisk}) may be due to different modeling of the disk evolution with time and different accretion torques.

For a normal neutron star at a period of $1091\ \rm s$, it is not sure whether they can lie above the pulsar death line or not (see discussions in section \ref{sect_death_line}). In our opinion, this is one difficulty for a normal neutron star. Future period-derivative observations of more sources may given us some information about the age of the neutron star. However, the theoretical period-derivative depends strongly on whether the disk is still active or not. Therefore, the period-derivative constraint may be model dependent. For young neutron+fallback disk system, the surround supernova remnant may be still visible (De Luca et al. 2006). Therefore, future observations of some supernova remnant associated with sources like GLEAM-X J1627 will support a young neutron star origin.

\begin{acknowledgments}
H.Tong would like to thank  Gao Yong for helpful discussions on precession of magnetars. This work is supported by National SKA Program of China (No. 2020SKA0120300) and NSFC (12133004).
\end{acknowledgments}









\end{document}